\documentclass[11pt]{article}
\usepackage{a4wide,amsmath,amssymb}
\usepackage[english]{babel}
\def\sym{\mathrm{Sym}\,}
\def\symt{\mathrm{SymTr}}
\def\Tr{\mathrm{Tr}}
\def\half{\frac{1}{2}}

\newfont{\bbbold}{msbm10 scaled \magstep1}

\def\bbW{\mbox{\bbbold W}}

\newfont{\goth}{eufm10 scaled \magstep1}

\def\a{\alpha}
\def\b{\beta}
\def\c{\gamma}
\def\d{\delta}
\def\e{\epsilon}
\def\f{\phi}

\def\l{\lambda}

\def\beq{\begin{equation}}\def\eeq{\end{equation}}
\def\beqa{\begin{eqnarray}}\def\eeqa{\end{eqnarray}}
\def\barr{\begin{array}}\def\earr{\end{array}}

\let\la=\label

\def\nn{\nonumber}
\def\bd{\begin{document}}
\def\ed{\end{document}}
\def\ba{\begin{array}}
\def\ea{\end{array}}
\def\bea{\begin{eqnarray}}
\def\eea{\end{eqnarray}}
\def\ft#1#2{{\textstyle{{\scriptstyle #1}\over {\scriptstyle #2}}}}
\def\fft#1#2{{#1 \over #2}}
\newcommand{\be}{\begin{equation}}
\newcommand{\ee}{\end{equation}}
\newcommand{\eq}[1]{(\ref{#1})}
\newcommand{\w}[1]{\\[0.#1cm]}
\def\eqs#1#2{(\ref{#1}-\ref{#2})}
\def\det{{\rm det\,}}
\def\tr{{\rm tr}}
\newcommand{\hoch}[1]{$\, ^{#1}$}
\newcommand{\tamphys}{\it\small Center for Theoretical Physics,
Texas A\&M University, College Station, TX 77843, USA}
\newcommand{\kings}
{\it\small Department of Mathematics, King's College, London, UK}
\newcommand{\uu}
{\it\small Theoretical Physics, Department of Physics and Astronomy, Uppsala, Sweden}
\newcommand{\hip}
{\it\small HIP-Helsinki Institute of Physics,
P.O. Box 64 FIN-00014 University of Helsinki, Suomi-Finland}
\newcommand{\stock}
{\it\small Department of Theoretical Physics, Stockholm, Sweden}
\newcommand{\pad}
{\it\small INFN, Sezione di Padova, via F. Marzolo 8, 35131 Padova, Italia}
\makeatletter
\renewcommand\theequation{\thesection.\arabic{equation}}
\@addtoreset{equation}{section} \makeatother

\newcommand{\auth}
{\large P.S. Howe\hoch{1}, U. Lindstr\"om\hoch{2} and L.
Wulff\hoch{3}}

\begin{document}

\hfill{KCL-TH-10-05}

\hfill{UUITP-13/10}

\vspace{20pt}

\begin{center}
{\Large{\bf $D=10$ supersymmetric Yang-Mills theory at $\a'^4$}}
\vspace{30pt}

\auth

\vspace{15pt}

\begin{itemize}
\item [$^1$] \kings \item [$^2$] \uu \item  [$^3$] \pad
\end{itemize}

\vspace{60pt}

{\bf Abstract}

\end{center}

The $\a'^2$ deformation of $D=10$ SYM is the natural generalisation of the $F^4$ term in the abelian Born-Infeld theory. It is shown that this deformation can be extended to $\a'^4$ in a way which is consistent with supersymmetry. The latter requires the presence of higher-derivative and commutator terms as well as the symmetrised trace of the Born-Infeld $\a'^4$ term.

\pagebreak \tableofcontents \setcounter{page}{1}


\section{Introduction}


We consider the problem of deforming ten-dimensional maximally supersymmetric Yang-Mills theory by higher-order corrections (controlled by $\alpha'$ in string theory) while preserving supersymmetry. The (unique) lowest-order correction, which occurs at order $\alpha'^2$, is in accord with the symmetrised trace Born-Infeld prescription \cite{Tseytlin:1997csa}, but it has been known for many years that this is not the whole story \cite{Kitazawa:1987xj}. Indeed, there is a term at $\a'^3$ that is not present in the abelian theory \cite{Bilal:2001hb,Refolli:2002gt,deRoo:2002ap,Medina:2002nk,Drummond:2003ex,Sevrin:2003vs,Chandia:2003sh},
a fact that can be understood from the point of view of spinorial cohomology \cite{Cederwall:2002df}
and also because it is the full superspace integral of the Konishi superfield \cite{Drummond:2003ex} which vanishes in the abelian limit. The purpose of the current paper is to extend the $\a'^2$ term to the next order, i.e. $\alpha'^4$.\footnote{The $\a'^3$ term is irrelevant for this and will not be considered further here.} This is already a non-trivial task and, although string theory considerations suggest that this should be possible, until now there has not been a proof that it can be done.\footnote{In \cite{Movshev:2009ba} it was claimed that the $\alpha'^2$-correction can be extended to any order in $\alpha'$ but no proof was given.} The only result at order $\alpha'^4$ in $D=10$ is the computation of the purely bosonic terms in the action carried out in \cite{Koerber:2002zb}, although some fully supersymmetric $\a'^4$ terms were written down for $D=4, N=4$ SYM in \cite{Drummond:2003ex}. The reason for the complications is that, although the $\alpha'^2$-correction is given by a symmetrised ordering of fields, so that the calculations at this order almost follow the abelian case, these will give rise to terms at order $\alpha'^4$ involving nested symmetric products of the form $\sym(AB\,\sym(CDE))$, where $A,B,C,D,E$ are SYM fields. Now $\sym(AB\,\sym(CDE))\neq\sym(ABCDE)$; in fact, they differ by terms that involve two and four commutators of $A,B,C,D,E$. This means that at this order the non-abelian nature of the gauge-group (which we take to be $U(k)$) really manifests itself.

The $\alpha'^2$ deformation of SYM is very special. It is the only single trace (i.e. arising at tree level in string theory) invariant that cannot be easily extended to all orders in $\alpha'$. The reason is that all other single trace invariants one can add to the action can be written as full superspace integrals \cite{Drummond:2003ex,Movshev:2009ba}, i.e. integrals over all 16 odd coordinates of the superspace. Such an expression is manifestly supersymmetric and, although adding a deformation like this changes the equations of motion (and the invariants are on-shell invariants, i.e. defined modulo lowest order equations of motion), these changes are accounted for order by order by other terms that are also full superspace integrals.

In the case of an abelian gauge-group the $\alpha'^2$-correction can be completed to all orders in $\alpha'$ and gives rise to supersymmetric Born-Infeld. It is the $p=9$ case of the action for a single D$p$-brane \cite{Cederwall:1996ri,Bergshoeff:1996tu,Aganagic:1996nn} in flat IIB superspace. In the non-abelian case it should give rise to something that might be called ''non-abelian supersymmetric Born-Infeld'' since it is the minimal invariant that includes abelian Born-Infeld. The name is perhaps a bit misleading as we will see that terms involving derivatives of the fields have to be included. In particular, terms that reduce to higher-derivative terms in the abelian case decouple and are therefore absent in the standard supersymmetric Born-Infeld approximation.

Let us note that the fact that higher derivative terms have to be included to get a supersymmetric action (as noted already in \cite{Koerber:2002zb}) is also the reason why many other attempts at constructing an action for coincident D-branes have not been so successful. In the abelian case the supersymmetric Born-Infeld action can be nicely derived using the superembedding formalism \cite{Howe:1996mx,Howe:1998tsa}, where kappa-symmetry is interpreted as worldvolume supersymmetry (for a review see \cite{Sorokin:1999jx}). This approach has been extended to the non-abelian case by including boundary fermions (representing Chan-Paton factors) for the string \cite{Howe:2005jz,Howe:2006rv,Howe:2007eb}. The boundary fermions are treated as classical until the end when they are quantised. This approach maintains manifest supersymmetry and kappa-symmetry up until the final step when the boundary fermions are quantised. In the bosonic case it can be used to obtain the action of Myers \cite{Myers:1999ps}, as was shown in \cite{Howe:2006rv}. Although this approach is pleasingly geometrical we show here that it is unfortunately incomplete since it does not include the higher-derivative terms needed for supersymmetry. This means that the (naive) quantisation of the boundary fermions spoils the supersymmetry. An interesting question, which we hope to return to in the near future, is whether the approach can be modified to fix this problem. For some recent alternative work on the superembedding approach to the problem see references \cite{Bandos:2009yp,Bandos:2009gk}.

The approach we will use is based on spinorial cohomology \cite{Cederwall:2001bt,Cederwall:2001td}  which is in turn related to the pure spinor formalism for supersymmetry in ten and eleven dimensions \cite{Howe:1991mf,Howe:1991bx,Tonin:1991ii,Berkovits:1991qg}; it was briefly outlined at $\a'^4$ in \cite{Howe:2008vb}. It makes use of the fact that $D=10$ SYM can be defined by the constraint on the $(0,2)$-component of the SYM field strength. The $\a'^2$ deformation induces higher-order terms and we shall work these out at order $\alpha'^4$. After setting up our conventions we describe the theory at lowest and first order before turning to the induced $\a'^4$ terms in section 5 where we are guided by results from the superembedding formalism. It is shown that further terms are necessary and these are described in the text. In section 6 we describe an alternative method which constructs the action directly using the ``ectoplasm'' formalism. We then state our conclusions in section 7 while there are three appendices giving more details of the main calculation.


\section{Conventions and tools}



\subsection{$\mathcal N=1$ superspace}


We take the supervielbein of flat $D=10$, $\mathcal N=1$ superspace to be
\begin{eqnarray}
E^a&=&dx^a-\frac{i}{2}d\theta\gamma^a \theta\nonumber\\
E^\alpha&=&d\theta^\alpha\,,
\end{eqnarray}
where $a=0,\ldots,9$ and $\alpha=1,\ldots,16$ are tangent space indices (we will often suppress the spinor indices as in the first line). The only non-vanishing components of the torsion are then
\begin{equation}
T_{\alpha\beta}{}^a=-i\gamma^a_{\alpha\beta}\,.
\end{equation}
A superspace $n$-form can be split invariantly into $(p,q)$-forms where $p+q=n$ and $p$ ($q$) denotes the even (odd) degree of the form,
\begin{equation}
\frac{1}{n!}E^{A_n}\cdots E^{A_1}\omega_{A_1\cdots A_n}=\sum_{p=0}^n\frac{1}{p!q!}E^{\alpha_q}\cdots E^{\alpha_1} E^{a_p}\cdots E^{a_1}\omega_{a_1\cdots a_p\alpha_1\cdots\alpha_q}\,.
\end{equation}
We will denote the $(p,q)$ component of $\omega$ as $\omega_{p,q}$.


\subsection{Conventions for super Yang-Mills}
The field strength of SYM is
\begin{equation}
F=dA-AA
\end{equation}
and the gauge-covariant derivative is
\begin{equation}
D=d-[\cdot\,,A]\,,
\end{equation}
so that
\begin{equation}
D^2=-[\cdot\,, F]\,.
\end{equation}
In components this means we get
\begin{equation}
[D_A,D_B]\Phi=[F_{AB},\Phi]\,,
\end{equation}
where the commutator becomes an anti-commutator if both $\Phi$ and the component of $F$ concerned are odd.

There is some redundancy in the definition of the gauge field which can be removed by imposing the following conventional constraints on the dimension zero and dimension one-half super Yang-Mills field strength
\begin{equation}
\gamma_a^{\alpha\beta}F_{\alpha\beta}=0
\end{equation}
and
\begin{equation}
F_{a\b}=-\frac{i}{2}(\gamma_a\lambda)_\b+\chi_{a\b}\,,
\label{eq:Falphab}
\end{equation}
where $\lambda$ is the spinor superfield and $\chi$ is gamma-traceless, i.e. $(\gamma^b)^{\alpha\beta}\chi_{b\b}=0$.

In components the Bianchi identity $DF=0$ reads
\begin{eqnarray}
D_{(\alpha}F_{\beta\gamma)}-i\gamma^c_{(\alpha\beta}F_{|c|\gamma)}&=&0\\
2D_{(\alpha}F_{\beta)c}+D_cF_{\alpha\beta}-i\gamma^d_{\alpha\beta}F_{dc}&=&0\\
D_\alpha F_{bc}+2D_{[b}F_{c]\alpha}&=&0\\
D_{[a}F_{bc]}&=&0\,.
\end{eqnarray}

The gamma-contraction of the lowest dimension Bianchi determines $\chi$ completely in terms of $F_{\alpha\beta}$
\begin{equation}
\chi_{a\beta}=-\frac{i}{10}\gamma_a^{\alpha\gamma}D_\alpha F_{\gamma\beta}\,.
\label{eq:chiabetadef}
\end{equation}
While the second (dimension 1) Bianchi identity gives
\begin{equation}
i(\gamma_cD_{(\alpha}\lambda)_{\beta)}=i\gamma^d_{\alpha\beta}F_{dc}-2D_{(\alpha}\chi_{c\beta)}-D_cF_{\alpha\beta}\,,
\label{eq:gammaDalphalambda}
\end{equation}
which determines $D_\alpha\lambda$ (recursively). Finally the third Bianchi identity gives
\begin{equation}
D_\alpha F_{bc}=i(\gamma_{[c}D_{b]}\lambda)_\alpha+2D_{[b}\chi_{c]\alpha }\,.
\label{eq:DalphaFab}
\end{equation}


\subsection{Superspace cohomology}


Superspace has some interesting notions of cohomology associated with it. This can be seen as follows. The exterior derivative can be split according to degree as \cite{Bonora:1986ix}
\begin{equation}
d=d_0+d_1+t_0
\end{equation}
(in curved superspace there would also be a $t_1$). The different components acts as follows: $d_0$ acts as an ordinary bosonic exterior derivative mapping $(p,q)$-forms to $(p+1,q)$-forms, $d_1$ acts as a purely spinorial exterior derivative mapping $(p,q)$-forms to $(p,q+1)$-forms and finally $t_0$ corresponds to the dimension zero torsion and maps $(p,q)$-forms to $(p-1,q+2)$-forms. Explicitly
\begin{equation}
t_0 E^a=T^a=-\frac{i}{2}E\gamma^aE\,.
\end{equation}
Now since $d^2=0$ it follows that
\begin{eqnarray}
t_0^2&=&0\\
d_1t_0+t_0d_1&=&0\\
d_1^2+t_0d_0+d_0t_0&=&0\,.
\end{eqnarray}
The first condition allows us to define $t_0$-cohomology \cite{Bonora:1986ix}. The cohomology groups $H^{p,q}_{t}$ consist of $t_0$-closed forms modulo $t_0$-exact terms. Throughout the paper we shall use the notation '$\sim$' to mean belonging to the same $t_0$ cohomology class. In components we have
\begin{equation}
X_{\alpha\beta\cdots}\sim Y_{\alpha\beta\cdots}\qquad\Leftrightarrow\qquad X_{\alpha\beta\cdots}=Y_{\alpha\beta\cdots}+\gamma^a_{(\alpha\beta}Z_{\cdots)a}\,,
\end{equation}
for some $Z$.

We can now go one step further and define what is known as spinorial cohomology \cite{Cederwall:2001dx,Howe:2003cy}
(it is essentially equivalent to pure spinor cohomology \cite{Berkovits:2008qw}). We see that on $t_0$ cohomology classes
\begin{equation}
d_1^2[\omega]=[t_0d_0\omega]=0\,,
\end{equation}
which means that we can define spinorial cohomology associated to the exterior derivative $d_s$,
\begin{equation}
d_s\omega=[d_1\omega]\ ,
\end{equation}
which maps elements of $H_t^{p,q}$ to elements of $H_t^{p,q+1}$ and which squares to zero. The spinorial cohomology groups are denoted $H_s^{p,q}$.

This construction can be straight-forwardly extended to the case of SYM (without $\alpha'$-corrections). The $t_0$-cohomology classes can be defined in exactly the same way. For the spinorial cohomology one now has
\begin{equation}
D_1^2\omega\sim [F_{0,2},\omega]\,.
\end{equation}
For pure SYM $F_{0,2}$ vanishes and it makes sense to define the spinorial cohomology group associated to $D_s$ where
\begin{equation}
D_s\omega\sim D_1\omega\,.
\end{equation}
When we work with expressions at order $\alpha'^4$ the Yang-Mills fields will be taken to satisfy the lowest-order equations of motion, i.e. those of pure SYM, so this notion of spinorial cohomology will be relevant for us.


\subsection{Deformations of SYM}


We start from the known $\alpha'^2$-correction to the constraint on the lowest-dimensional component of the SYM field strength, $F_{0,2}=F_{0,2}^1+\mathcal O(\alpha'^4)$. Using this we can compute $D_1\lambda$ and $D_1F_{2,0}$ up to first order ($\alpha'^2$). This then allows us to compute $D_1F_{0,2}^1$ to \emph{second} order ($\alpha'^4$). The Bianchi identity for the dimension zero field strength says that $D_sF_{0,2}\sim0$. Writing $F_{0,2}=F_{0,2}^1+F_{0,2}^2+\mathcal O(\alpha'^6)$ we have
\begin{equation}
D_sF_{0,2}^1+D_sF_{0,2}^2\sim\mathcal O(\alpha'^6)\,.
\end{equation}
Since the first term is known, the problem is to find $F_{0,2}^2$ such that this equation is satisfied (using the lowest order expressions for $D_1\lambda$ and $D_1F_{2,0}$ as well as equations of motion). In other words we have to show that $(D_sF^1_{0,2})^2$ is trivial in $H_s^{0,3}$. A priori there is no guarantee that this cohomological problem has a solution, i.e. it could be that $(D_sF_{0,2}^1)^2$ is not exact. We will see that in our case there is no obstruction and a solution for $F_{0,2}^2$ can be found (as suggested by string theory). The solution will turn out to be quite non-trivial however and in particular will involve higher-derivative (and commutator) terms.

We will now describe SYM at lowest order ($\alpha'^0$).


\section{Lowest order or pure super Yang-Mills}


At lowest order super Yang-Mills is defined by
\begin{equation}
F_{\alpha\beta}^0=0\,,
\end{equation}
or in form notation $F_{0,2}^0=0$. Since, using (\ref{eq:chiabetadef}), this implies $\chi^0=0$ we immediately see from (\ref{eq:gammaDalphalambda}) that
\begin{equation}
\label{eq:Dalphalambda0}
(D_\alpha\lambda^\beta)^0=\frac{1}{2}F_{ab}(\gamma^{ab})_\alpha{}^\beta
\end{equation}
and from (\ref{eq:DalphaFab}) that
\begin{equation}
\label{eq:DalphaFab0}
(D_\alpha F_{bc})^0=i(\gamma_{[c}D_{b]}\lambda)_\alpha\,.
\end{equation}
We will also need the equations of motion which can be obtained by using the above expressions and the identity
\begin{eqnarray}
0&=&D_{(\alpha}D_{\beta)}\lambda^\gamma-\frac{i}{2}\gamma^c_{\alpha\beta}D_c\lambda^\gamma
\simeq
\frac{1}{2}D_{(\alpha}F_{ab}(\gamma^{ab})_{\beta)}{}^\gamma
-\frac{i}{2}\gamma^c_{\alpha\beta}D_c\lambda^\gamma
\nonumber\\
&\simeq&
\frac{i}{2}(\gamma_bD_a\lambda)_{(\alpha}(\gamma^{ab})_{\beta)}{}^\gamma
-\frac{i}{2}\gamma^c_{\alpha\beta}D_c\lambda^\gamma\,,
\end{eqnarray}
which implies the Dirac equation for the fermion
\begin{equation}
i(\gamma^aD_a\lambda)^0=0\,.
\label{eq:Diraceqn}
\end{equation}
The bosonic equation is easily obtained by taking a spinor derivative of the equation for the fermion. We find
\begin{equation}
0\simeq(\gamma^aD_{(\beta}D_a\lambda)_{\alpha)}
=
(\gamma^a\{F_{(\beta a},\lambda\})_{\alpha)}
+(\gamma^aD_aD_{(\beta}\lambda)_{\alpha)}
\simeq
\gamma^b_{\alpha\beta}(D_aF_b{}^a-\frac{i}{4}\{\lambda,\gamma_b\lambda\})
\end{equation}
so that
\begin{equation}
(D^bF_{ab})^0=\frac{i}{4}\{\lambda,\gamma_a\lambda\}\,.
\label{eq:BIeqn}
\end{equation}
We now turn to the first supersymmetric correction to SYM which occurs at order $\alpha'^2$.


\section{First correction (order $\alpha'^2$)}


The first correction to the dimension zero component of $F$ was given in \cite{Gates:1986is,Bergshoeff:1986jm} while  cohomological methods were used to show that it is unique up to field-redefinitions in \cite{Cederwall:2001bt,Cederwall:2001td}. It is given by
\begin{eqnarray}
F_{\alpha\beta}^1&=&\frac{1}{4\cdot2\cdot16\cdot5!}\gamma^{abcde}_{\alpha\beta}\sym(\lambda\gamma^f\gamma_{abcde}\gamma^g\lambda F_{gf})
\nonumber\\
&=&
\frac{1}{4}\sym((\gamma^a\lambda)_\alpha(\gamma^b\lambda)_\beta F_{ba})
-\frac{1}{4\cdot16}\gamma^c_{\alpha\beta}\sym(\lambda\gamma_c{}^{ab}\lambda\,F_{ab})\,,
\label{eq:Falphabeta1}
\end{eqnarray}
where $\sym$ stands for symmetrised ordering of the fields (see Appendix A). The second expression will be more convenient for some of our calculations since only the first term is non-trivial in $t_0$-cohomology .
In order to find $\chi^1$, the first correction to $\chi$ (defined in (\ref{eq:Falphab})), we compute, using the lowest order expressions for $D_\alpha F_{ab}$ and $D_\alpha\lambda$ in (\ref{eq:DalphaFab0}) and (\ref{eq:Dalphalambda0})
\begin{eqnarray}
D_\alpha F_{\beta\gamma}^1
&=&
\frac{1}{2\cdot32\cdot5!}\gamma^{abcde}_{\beta\gamma}\sym(D_\alpha\lambda\gamma^f\gamma_{abcde}\gamma^g\lambda F_{gf})
+\frac{1}{4\cdot32\cdot5!}\gamma^{abcde}_{\beta\gamma}\sym(\lambda\gamma^f\gamma_{abcde}\gamma^g\lambda D_\alpha F_{gf})
\nonumber\\
&\simeq&
\frac{1}{8\cdot16\cdot5!}\gamma^{abcde}_{\beta\gamma}\sym((\gamma^{hi}\gamma^f\gamma_{abcde}\gamma^g\lambda)_\alpha F_{hi}F_{gf})
+\frac{i}{4}\sym((\gamma^a\lambda)_{(\beta}(\gamma^b\lambda)_{\gamma)}(\gamma_aD_b\lambda)_\alpha)
\nonumber\\
&&{}
+\frac{i}{4\cdot16}\gamma^c_{\beta\gamma}\sym(\lambda\gamma_c{}^{ab}\lambda\,(\gamma_aD_b\lambda)_\alpha)\,,
\end{eqnarray}
where we've replaced the $\gamma^{(5)}$-part of the last term by the symmetric part minus the $\gamma^{(1)}$-part to simplify the calculation. Using this we get
\begin{eqnarray}
\chi^1_{\beta k}&=&\frac{i}{10}\gamma_k^{\alpha\gamma}(D_\alpha F_{\gamma\beta})^1
\nonumber\\
&=&
\frac{i}{40\cdot32\cdot5!}\sym((\gamma^{abcde}\gamma_k\gamma^{hi}\gamma^f\gamma_{abcde}\gamma^g\lambda)_\beta F_{hi}F_{gf})
-\frac{1}{80}\sym((\gamma^a\lambda)_\beta\,\lambda\gamma^b\gamma_k\gamma_aD_b\lambda)
\nonumber\\
&&{}
-\frac{1}{10}\sym((\gamma^a\lambda)_\beta\,\lambda\gamma_kD_a\lambda)
-\frac{1}{40\cdot16}\sym(\lambda\gamma_c{}^{ab}\lambda\,(\gamma^c\gamma_k\gamma_aD_b\lambda)_\beta)\,.
\end{eqnarray}
Using the gamma-matrix identities
\begin{eqnarray}
\gamma^{abcde}\gamma_{abcde}&=&\frac{10!}{5!}\nonumber\\
\gamma^{abcde}\gamma^{(2)}\gamma_{abcde}&=&-4\frac{7!}{3!}\gamma^{(2)}\nonumber\\
\gamma^{abcde}\gamma^{(4)}\gamma_{abcde}&=&2\cdot6!\gamma^{(4)}
\end{eqnarray}
together with the lowest order equation of motion for $\lambda$ we get
\begin{eqnarray}
\chi^1_{\beta k}&=&
-\frac{3i}{32}\sym((\gamma^{abc}\lambda)_\beta F_{ab}F_{ck})
-\frac{7i}{16}\sym((\gamma^a\lambda)_\beta F_{ab}F^b{}_k)
-\frac{1}{32}\sym((\gamma^a\lambda)_\beta\,\lambda\gamma_aD_k\lambda)
\nonumber\\
&&{}
-\frac{1}{16}\sym((\gamma^a\lambda)_\beta\,\lambda\gamma_kD_a\lambda)
-\frac{3i}{320}\sym((\gamma_k\gamma^{abcd}\lambda)_\beta F_{ab}F_{cd})
+\frac{7i}{160}\sym((\gamma_k\lambda)_\beta F_{ab}F^{ba})
\nonumber\\
&&{}
+\frac{1}{320}\sym((\gamma_k\gamma^{ab}\lambda)_\beta\,\lambda\gamma_aD_b\lambda)\,.
\label{eq:chi1}
\end{eqnarray}
The last three terms ensure that $\chi$ is indeed gamma-traceless. Knowing $\chi^1$, the first correction to $F_{a\beta}$, allows us to compute the first correction to $D_\alpha\lambda$ and $D_\alpha F_{ab}$. In particular we have
\begin{equation}
i(\gamma_cD_{(\alpha}\lambda)_{\beta)}=i\gamma^d_{\alpha\beta}F_{dc}-2D_{(\alpha}\chi_{\beta)c}-D_cF_{\alpha\beta}\,,
\end{equation}
so we see that we need to compute $D_{(\alpha}\chi^1_{\beta)k}$. In fact it turns out that for our calculations we will only need $D_s\chi$, i.e. we can work in $t_0$-cohomology. Doing this we find that the expression consists of four types of terms, schematically
\begin{equation}
(D_{(\alpha}\chi_{\beta)k})^1\sim(F^3)+(\lambda\lambda DF)+(\lambda\{\lambda,\lambda\})+(\lambda D\lambda F)\,.
\end{equation}
The $F^3$-terms are the easiest. They come from the first and fifth term in the expression for $\chi^1$ and a short calculation gives
\begin{eqnarray}
(F^3)&\sim&\frac{3i}{160}\gamma^{abcde}_{\alpha\beta}\sym(F_{ab}F_{cd}F_{ek})\,.
\end{eqnarray}
Terms with $DF$ and $\{\lambda,\lambda\}$ come from the third, fourth and last term. We get using the lowest order equation of motion and the Bianchi identity for $F_{ab}$
\begin{eqnarray}
(\lambda\lambda DF)+(\lambda\{\lambda,\lambda\})&\sim&
-\frac{3i}{160}\sym((\gamma^a\lambda)_{(\beta}\,\{(\gamma_k\lambda)_{\alpha)},\lambda\}\gamma_a\lambda)
\nonumber\\
&&{}
-\frac{1}{40}\sym((\gamma^a\lambda)_{(\beta}\,(\gamma_k\gamma^{cd}\lambda)_{\alpha)} D_aF_{cd})
-\frac{1}{8}\sym((\gamma^a\lambda)_{(\beta}\,(\gamma^b\lambda)_{\alpha)} D_kF_{ba})\,.
\nonumber\\
\end{eqnarray}
Finally we have the terms
\begin{eqnarray}
(\lambda D\lambda F)&\sim&
-\frac{1}{640}(\gamma_k{}^{abcd})_{\alpha\beta}\sym(\lambda\gamma_aD_b\lambda F_{cd})
-\frac{3}{320}\sym((\gamma^{abc}\lambda)_{(\beta}(\gamma_kD_c\lambda)_{\alpha)}F_{ab})
\nonumber\\
&&{}
-\frac{7}{160}\sym((\gamma_k\gamma^{ab}\lambda)_{(\beta}(\gamma^cD_a\lambda)_{\alpha)}F_{bc})
+\frac{9}{320}\sym((\gamma^a\lambda)_{(\beta}(\gamma_k\gamma^{cd}D_a\lambda)_{\alpha)} F_{cd})
\nonumber\\
&&{}
+\frac{13}{160}\sym((\gamma_k\lambda)_{(\beta}(\gamma^aD^b\lambda)_{\alpha)} F_{ab})
+\frac{7}{32}\sym((\gamma^a\lambda)_{(\beta}(\gamma_kD^b\lambda)_{\alpha)} F_{ba})
\nonumber\\
&&{}
-\frac{1}{4}\sym((\gamma^a\lambda)_{(\beta}(\gamma^bD_k\lambda)_{\alpha)}F_{ba})\,.
\end{eqnarray}
For the combination appearing in $D_\alpha\lambda$ the last term from each expression above cancels the contributions from $D_kF_{\alpha\beta}^1$ and we have
\begin{eqnarray}
\lefteqn{(D_{(\alpha}\chi_{\beta)k}+\frac{1}{2}D_kF_{\alpha\beta})^1}
\nonumber\\
&\sim&
\frac{3i}{160}\gamma^{abcde}_{\alpha\beta}\sym(F_{ab}F_{cd}F_{ek})
-\frac{3i}{160}\sym((\gamma^a\lambda)_{(\beta}\,\{(\gamma_k\lambda)_{\alpha)},\lambda\}\gamma_a\lambda)
\nonumber\\
&&{}
-\frac{1}{40}\sym((\gamma^a\lambda)_{(\beta}\,(\gamma_k\gamma^{cd}\lambda)_{\alpha)} D_aF_{cd})
-\frac{1}{640}(\gamma_k{}^{abcd})_{\alpha\beta}\sym(\lambda\gamma_aD_b\lambda F_{cd})
\nonumber\\
&&{}
-\frac{3}{320}\sym((\gamma^{abc}\lambda)_{(\beta}(\gamma_kD_c\lambda)_{\alpha)}F_{ab})
-\frac{7}{160}\sym((\gamma_k\gamma^{ab}\lambda)_{(\beta}(\gamma^cD_a\lambda)_{\alpha)}F_{bc})
\nonumber\\
&&{}
+\frac{9}{320}\sym((\gamma^a\lambda)_{(\beta}(\gamma_k\gamma^{cd}D_a\lambda)_{\alpha)} F_{cd})
+\frac{13}{160}\sym((\gamma_k\lambda)_{(\beta}(\gamma^aD^b\lambda)_{\alpha)} F_{ab})
\nonumber\\
&&{}
+\frac{7}{32}\sym((\gamma^a\lambda)_{(\beta}(\gamma_kD^b\lambda)_{\alpha)} F_{ba})\,.
\label{eq:Dalphachi}
\end{eqnarray}
Using these expressions one can compute explicitly the $\alpha'^2$-corrections to $D_\alpha\lambda$ and $D_\alpha F_{ab}$ (although one would have to reinstate the $t_0$-exact terms). However we will only need the expressions computed here to address the problem of constructing $F_{\alpha\beta}$ at the next order, i.e. $\alpha'^4$.


\section{Induced terms at second order ($\alpha'^4$)}

We now want to see what terms in $F_{\alpha\beta}$ are generated at the next order, \emph{i.e.} $\alpha'^4$, by requiring the $\alpha'^2$ correction to be consistent with the Bianchi identities. From (\ref{eq:Falphabeta1}) we have
\begin{eqnarray}
D_{(\alpha}F^1_{\beta\gamma)}&\sim&
\frac{1}{2}\sym((\gamma^aD_{(\alpha}\lambda)_\beta(\gamma^b\lambda)_{\gamma)} F_{ba})
+\frac{1}{4}\sym((\gamma^a\lambda)_{(\beta}(\gamma^b\lambda)_\gamma D_{\alpha)}F_{ba})\,.
\end{eqnarray}
Note that $D_\alpha\lambda$ and $D_\alpha F_{ba}$ are treated as a single object in the symmetrization even though they are composite fields beyond lowest order in $\alpha'$. We are only interested in terms which cannot be compensated for by $\chi^2$, \emph{i.e.} terms which are non-trivial elements of the $t_0$-cohomology, hence the $\sim$ instead of equality. Using the Bianchi identities, in the form (\ref{eq:gammaDalphalambda}) and (\ref{eq:DalphaFab}), we get at second order
\begin{eqnarray}
(D_{(\alpha}F^1_{\beta\gamma)})^2&\sim&
i\sym((\gamma_a\lambda)_{(\alpha}F^{ab}[D_\beta\chi_{\gamma)b}+\frac{1}{2}D_bF_{\beta\gamma)}]^1)
+\frac{1}{2}\sym((\gamma^a\lambda)_{(\alpha}(\gamma^b\lambda)_\beta D_b\chi^1_{\gamma)a})\,.
%
%
\label{eq:DalphaF1}
\end{eqnarray}
The quantity in brackets is precisely what we computed in (\ref{eq:Dalphachi}), while $\chi^1$ was computed in (\ref{eq:chi1}). In fact the terms in $D_b\chi_a$ of the form $\gamma_a\cdot\ldots$ give rise to $t_0$-exact terms, so these terms can be dropped. The only terms in $\chi^1$ that contribute to the second term are therefore
\begin{eqnarray}
\chi'^1_{\beta k}&=&
-\frac{3i}{32}\sym((\gamma^{abc}\lambda)_\beta F_{ab}F_{ck})
-\frac{7i}{16}\sym((\gamma^a\lambda)_\beta F_{ab}F^b{}_k)
-\frac{1}{32}\sym((\gamma^a\lambda)_\beta\,\lambda\gamma_aD_k\lambda)
\nonumber\\
&&{}
-\frac{1}{16}\sym((\gamma^a\lambda)_\beta\,\lambda\gamma_kD_a\lambda)\,.
\label{eq:chiprime}
\end{eqnarray}

We can now compute $D_{(\alpha}F^1_{\beta\gamma)}$ or, in more compact form notation, $D_sF_{0,2}^1$ at second order ($\alpha'^4$).
Using equations (\ref{eq:Dalphachi}) and (\ref{eq:chiprime}) in (\ref{eq:DalphaF1}) a short calculation gives
\begin{eqnarray}
(D_sF^1_{0,2})^2&\sim&
\frac{3}{160}A_1^{(15(234))}
-\frac{3}{64}B_1^{(12(345))}
+\frac{1}{40}B_1^{(15(234))}
-\frac{3}{64}B_2^{(12(345))}
-\frac{1}{8}B_3^{(12(345))}
\nonumber\\
&&{}
-\frac{1}{10}B_3^{(15(234))}
+\frac{1}{8}B_3^{(23(145))}
-\frac{3}{64}C_1^{(12(345))}
+\frac{1}{32}C_1^{(14(523))}
-\frac{1}{32}C_1^{(24(513))}
\nonumber\\
&&{}
-\frac{1}{32}C_2^{(12(345))}
-\frac{1}{160}D_2^{(14(235))}
+\frac{1}{20}D_3^{(14(235))}
-\frac{3}{64}D_4^{(12(345))}
-\frac{1}{40}D_4^{(15(234))}
\nonumber\\
&&{}
+\frac{1}{20}D_5^{(15(234))}
+\frac{1}{5}D_5^{(24(135))}
+\frac{3}{40}D_6^{(14(235))}
-\frac{1}{8}D_7^{(12(345))}
+\frac{3}{20}D_7^{(15(234))}
\nonumber\\
&&{}
-\frac{1}{32}E^{(12(345))}
+\frac{3}{160}K^{(16(3[24]5))}
-\frac{1}{128}K^{(12(34[56]))}\,.
\label{eq:DsF1}
\end{eqnarray}
Let us explain the notation used here. It is convenient to use a notation where the structure of a term as far as the fields it contains, $\lambda$,  $F_{ab}$ and covariant derivatives thereof, and the gamma-matrix structure is separated from the gauge structure, \emph{i.e.} the ordering of the fields. The letters $A$--$K$ denote terms with different structure of fields and covariant derivatives and the subscript denotes terms with different gamma-matrix structure. The relevant terms are defined in Appendix B. Finally the superscript denotes the gauge structure \emph{i.e.} the ordering of the fields. The fields are labeled $1,2,3,\ldots$ according to the order in which they appear in the definition of a given term in Appendix B, for example $B_1^{13245}$ means $B_1$ as defined in Appendix B but with the second and third factor interchanged. Care must be taken to include a minus sign when two fermionic factors are interchanged. Finally (graded) commutators are denoted for example $[12]$ in superscripts and (graded) symmetrization for example $(123)$.

In order to construct super Yang-Mills at this order one should find $F_{\alpha\beta}^2$, the $\alpha'^4$-correction to $F_{\alpha\beta}$, such that
\begin{equation}
D_sF^1_{0,2}+D_sF^2_{0,2}\sim\mathcal O(\alpha'^6)\,.
\end{equation}
Our task is to determine $F_{0,2}^2$. To get an idea of the terms $F_{0,2}^2$ should contain we will first use the results of the generalised superembedding formalism. This will turn out not to give us the full answer but it will get us some of the way.

\subsection{Comparison to the superembedding approach}
The superembedding formalism, generalised to the non-abelian case by including boundary fermions (representing Chan-Paton degrees of freedom) for the string, can be used to derive super Yang-Mills including $\alpha'$-corrections \cite{Howe:2005jz,Howe:2006rv,Howe:2007eb}. Due to the semi-classical treatment of the boundary fermions the ordering of the fields in the case of a non-abelian gauge-group is however not determined. The ''natural'' ordering, \emph{i.e.} symmetrised ordering, is correct at lowest order ($\alpha'^2$) but fails at the next order as we will demonstrate.

The expressions that follow from this generalised superembedding approach are \cite{Howe:2005jz}\footnote{These were also derived from the pure spinor string in \cite{Berkovits:2002ag}.}

\begin{eqnarray}
f_{a\beta}&=&i(\gamma_a\Lambda)_\beta+\frac{1}{6}D_a\Lambda\gamma^b\Lambda\,(\gamma_b\Lambda)_\beta-\frac{i}{2}(h\gamma^b\Lambda)_\beta f_{ab}\\
f_{\alpha\beta}&=&\frac{1}{3}D_{(\alpha}\Lambda\gamma^a\Lambda\,(\gamma_a\Lambda)_{\beta)}
-\frac{1}{4}(h\gamma^a\Lambda)_\alpha(h\gamma^b\Lambda)_\beta f_{ab}
+\frac{1}{36}\Lambda\gamma^a(\Lambda,\Lambda)\gamma^b\Lambda\,(\gamma_a\Lambda)_\alpha(\gamma_b\Lambda)_\beta\,.
\label{eq:falphabeta}
\end{eqnarray}
The super Yang-Mills field strength and the spinor field are denoted $f$ and $\Lambda$ here since they differ by a field redefinition (that we will determine below) from the ones considered in the previous sections. Note that these expressions are exact in the sense that they incorporate a complete set of consistent $\alpha'$-corrections (modulo the subtlety with the ordering in the non-abelian case mentioned above). The expression $(\Lambda,\Lambda)$ denotes a Poisson bracket with respect to the boundary fermions and is to be interpreted as a (anti-) commutator upon their quantization. The spinorial derivative of $\Lambda$ appearing in the above equation is given by
\begin{equation}
D_\alpha\Lambda^\beta=h_\alpha{}^\gamma(\delta_\gamma^\beta-\frac{i}{2}(\gamma^a\Lambda)_\gamma D_a\Lambda^\beta)
-\frac{1}{6}(\gamma^a\Lambda)_\alpha\,\Lambda\gamma_a(\Lambda,\Lambda^\beta)\,,
\label{eq:DalphaLambda}
\end{equation}
while $h_\alpha{}^\beta$ can be determined from the dimension 1 Bianchi identity, using
\begin{equation}
D_\alpha f_{ab}=2i(\gamma_{[a}D_{b]}\Lambda)_\alpha+\mathcal O(\alpha'^2)\,.
\end{equation}
This gives the following expression for $h$ up to order $\alpha'^2$
\begin{eqnarray}
h_\alpha{}^\beta&=&-\frac{1}{4}(\gamma^{ab})_\alpha{}^\beta\Big(f_{ab}(1+\frac{1}{8}f_{cd}f^{cd})+\frac{1}{2}f^3_{ab}+iD_a\Lambda\gamma^c\Lambda\,f_{cb}\Big)
+\frac{1}{3\cdot64}(\gamma^{abcdef})_\alpha{}^\beta\,f_{ab}f_{cd}f_{ef}
\nonumber\\
&&{}+\mathcal O(\alpha'^4)\,.
\label{eq:halphabeta}
\end{eqnarray}
It is possible to give an exact (all orders in $\alpha'$) expression for $h$ but we will not need it here.

We will now determine the field redefinition that relates the conventionally defined fields $F$ and $\lambda$ to the ones appearing in the superembedding equations. Let $a$ be the potential corresponding to $f$, \emph{i.e.} $f=da-a\wedge a$, while $F=dA-A\wedge A$. Next we write $a_b=A_b+a_b'$ and require the conventional constraint $\gamma_a^{\alpha\beta}F_{\alpha\beta}=\gamma_a^{\alpha\beta}(2D_\alpha A_\beta+i\gamma^b_{\alpha\beta}A_b+[A_\alpha,A_\beta])=0$. Using $A_\alpha=a_\alpha$ together with the expression for $f_{\alpha\beta}$ to first order ($\alpha'^2$)
\begin{equation}
f_{\alpha\beta}^1=\frac{1}{3}(\gamma^a\Lambda)(\gamma^b\Lambda)f_{ba}
\end{equation}
we get
\begin{equation}
a_b'=-\frac{i}{48}\Lambda\gamma_b{}^{cd}\Lambda\,f_{cd}+\mathcal O(\alpha'^4)\,.
\end{equation}

The spinor field $\lambda$ is conventionally defined so that $(\gamma^a)^{\alpha\beta}F_{a\beta}=-5i\lambda$. Using the expression for $a_b'$ as well as the expression for $f_{a\beta}$ this condition gives
\begin{eqnarray}
\Lambda=-\frac{1}{2}\lambda-\frac{1}{120}\Lambda\,f_{ab}f^{ab}+\frac{i}{30}(\gamma^{ab}\Lambda)\,\Lambda\gamma_aD_b\Lambda
-\frac{1}{60}(\gamma^{abcd}\Lambda)\,f_{ab}f_{cd}+\mathcal O(\alpha'^4)\,.
\end{eqnarray}
And finally we have, using again the expression for $a_b'$,
\begin{equation}
F_{ab}=f_{ab}-\frac{i}{12}\Lambda\gamma_{[a}{}^{cd}D_{b]}\Lambda\,f_{cd}-\frac{i}{24}\Lambda\gamma_{[a}{}^{cd}\Lambda\,D_{b]}f_{cd}
+\mathcal O(\alpha'^4)\,.
\end{equation}
We will now consider the terms in $F_{0,2}^2$ coming from the superembedding approach (with symmetrised ordering).


\subsection{Terms at order $\alpha'^4$ coming from the superembedding approach}

Motivated by the analysis of the superembedding approach in the previous section we make the field redefinitions
\begin{eqnarray}
\Lambda&=&-\frac{1}{2}\lambda+\frac{1}{240}\sym(\lambda\,F_{ab}F^{ab})-\frac{i}{240}\sym((\gamma^{ab}\lambda)\,\lambda\gamma_aD_b\lambda)
+\frac{1}{120}\sym((\gamma^{abcd}\lambda)\,F_{ab}F_{cd})+\mathcal O(\alpha'^4)
\nonumber\\
f_{ab}&=&F_{ab}+\frac{i}{48}\sym(\lambda\gamma_{[a}{}^{cd}D_{b]}\lambda\,F_{cd})+\frac{i}{96}\sym(\lambda\gamma_{[a}{}^{cd}\lambda\,D_{b]}F_{cd})
+\mathcal O(\alpha'^4)\,.
\end{eqnarray}
Note that although the ordering is not determined by the generalised superembedding approach (when the boundary fermions are treated classically) it is natural to take the symmetrised ordering, indeed we know that this is the ordering that appears at order $\alpha'^2$.

Performing these field redefinitions the lowest order ($\alpha'^2$) terms in $f_{\alpha\beta}$ in (\ref{eq:falphabeta}) gives rise to additional terms at order $\alpha'^4$. We find
\begin{equation}
-\frac{1}{3}\sym((\gamma^a\Lambda)\,(\gamma^b\Lambda)\,f_{ab})\sim-\frac{1}{12}\sym((\gamma^a\lambda)\,(\gamma^b\lambda)\,F_{ab})+F^{(redef)}+\mathcal O(\alpha'^6)\,,
\end{equation}
where
\begin{eqnarray}
F^{(redef)}&\sim&
\frac{1}{9}\Big(
\frac{1}{80}Y_1^{(13(245))}
-\frac{1}{40}Y_5^{(23(145))}
+\frac{1}{32}Y_7^{(12(345))}
+\frac{1}{64}Y_{11}^{(12(345))}
-\frac{1}{80}Y_{12}^{(15(234))}
\nonumber\\
&&{}
-\frac{1}{128}Y_{13}^{(12(345))}
\Big)\,,
\label{eq:Fredef}
\end{eqnarray}
where the terms are defined in Appendix B.

In addition to this we have the $\alpha'^4$ terms in $F_{0,2}$ coming from the $\alpha'^4$ terms in $f_{\alpha\beta}$ in (\ref{eq:falphabeta}). From the first term in (\ref{eq:falphabeta}) we get
\begin{equation}
F^{(SE1)}=-\frac{1}{24}\sym((D_{(\alpha}\lambda)^1\gamma^a\lambda\,(\gamma_a\lambda)_{\beta)})\,.
\end{equation}
Using the superembedding expression for $D_\alpha\Lambda$, (\ref{eq:DalphaLambda}) and (\ref{eq:halphabeta}) and using the symmetrised ordering we find
\begin{eqnarray}
(D_\alpha\lambda^\beta)^1&=&
\frac{1}{2}(\gamma^{ab})_\alpha{}^\beta\sym\big(\frac{1}{8}F_{ab}F_{cd}F^{cd}+\frac{1}{2}F_{ac}F^{cd}F_{db}+\frac{i}{4}D_a\lambda\gamma^c\lambda\,F_{cb}\big)
\nonumber\\
&&{}
-\frac{1}{3\cdot32}(\gamma^{abcdef})_\alpha{}^\beta\,\sym(F_{ab}F_{cd}F_{ef})
-\frac{i}{16}\sym((\gamma^{bc}\gamma^a\lambda)_\alpha\,D_a\lambda^\beta\,F_{bc})
\nonumber\\
&&{}
+\frac{1}{48}\sym((\gamma^a\lambda)_\alpha\,\lambda\gamma_a\{\lambda,\lambda^\beta\})\,.
\end{eqnarray}
(This could of course be verified directly, without using the superembedding approach, but we will not need to do this here).

We therefore have
\begin{eqnarray}
F^{(SE1)}&=&
-\frac{1}{24}\Big(
\frac{1}{4}Y_1^{(12(345))}
+2Y_2^{(12(345))}
-\frac{1}{2}Y_3^{(12(345))}
-\frac{1}{8}Y_5^{(12(345))}
+\frac{1}{2}Y_7^{(12(345))}
\nonumber\\
&&{}
-\frac{1}{16}Y_9^{(13(452))}
+\frac{1}{48}Z^{(13([45]62))}
\Big)\,.
\label{eq:FSE1}
\end{eqnarray}

Finally we have the last two terms in the superembedding expression for $f_{\alpha\beta}$, (\ref{eq:falphabeta}), and which start at order $\alpha'^4$. Taking the symmetrised ordering they are
\begin{equation}
F^{(SE2)}=-\frac{1}{256}Y_4^{(12345)}+\frac{1}{9\cdot256}Z^{(123[45]6)}\,.
\label{eq:FSE2}
\end{equation}

Taking into account the terms coming from the superembedding approach, $F^{SE}=F^{redef}+F^{SE1}+F^{SE2}$, and using the spinorial derivatives of the terms and identities given in Appendix B and the expression for $D_sF^1_{0,2}$ in (\ref{eq:DsF1}) we find
\begin{eqnarray}
(D_sF_{0,2}^1)^2+9D_sF^{SE}&\sim&
\frac{9}{64}(A_1^{(15(234))}-A_1^{(12345)}-D_2^{(14(235))}+D_2^{(12345)})
\nonumber\\
&&{}
\hspace{-1cm}+\frac{1}{64}(3K^{(16(34[52]))}+2K^{(34([51]26))}+K^{(34([56]12))}-K^{(136[24]5)}+K^{(132[64]5)})\,,
\nonumber\\
\label{eq:DFplusFSE}
\end{eqnarray}
where the factor $9$ multiplying the superembedding terms is due to the fact that the $\alpha'^2$-correction has an extra factor of $\frac{1}{3}$ in the superembedding expression (\ref{eq:falphabeta}).

We see that, except when the gauge-group is abelian, the superembedding approach (with symmetrised ordering) gives \emph{almost} the complete answer at order $\alpha'^4$ but not quite (unlike at order $\alpha'^2$ where it gives the complete answer). We now turn to the problem of finding the terms not captured by the superembedding formalism. We will see that they are related to what amounts to higher derivative corrections in the abelian case.

The key idea to understanding how to cancel these remaining terms is to rewrite the terms with nested symmetrizations, e.g. $A_1^{(15(234))}$, which have the form
\begin{equation}
\sym(AB\,\sym(CDE))\,,
\end{equation}
where $A,B,C,D$ and $E$ are fields ($F$ or $\lambda$ or derivatives thereof), using the symmetrised ordering identities given in Appendix A as
\begin{equation}
\sym(AB\,\sym(CDE))=\sym(ABCDE)+\mbox{Two-commutator terms}+\mbox{Four-commutators terms}\,.
\end{equation}
The terms without additional commutators then cancel in (\ref{eq:DFplusFSE}).

This leaves us with terms with two (or three from the $K$-terms) and four (or five) commutators that can not be accounted for by the superembedding formalism. These should instead be related to higher derivative terms. The need to include what corresponds to higher derivative terms in the abelian case, in particular the $\partial^4F^4$-terms in the action at order $\alpha'^4$ was already noticed by Koerber and Sevrin in \cite{Koerber:2002zb} when they derived the bosonic terms in the action at this order.

The terms with two commutators on the RHS in (\ref{eq:DFplusFSE}) are (see Appendix A)
\begin{eqnarray}
\lefteqn{\frac{9}{64}(A_1^{(15(234))}-D_2^{(14(235))}-A_1^{(12345)}+D_2^{(12345)})_2}
\nonumber\\
&=&
\frac{3}{256}\Big(
2A_1^{([14][52]3)}+2A_1^{([12][54]3)}+2A_1^{([12][53]4)}
+2A_1^{([[52]4]31)}+2A_1^{([[54]2]31)}+2A_1^{([[52]3]41)}
\nonumber\\
&&{}
+2A_1^{([[12]4]35)}+2A_1^{([[14]2]35)}+2A_1^{([[12]3]45)}
+D_2^{([13][45]2)}+D_2^{([15][43]2)}
\nonumber\\
&&{}
-D_2^{([15][42]3)}-D_2^{([12][45]3)}
+D_2^{([13][42]5)}-D_2^{([12][43]5)}
+D_2^{([[45]3]21)}+D_2^{([[43]5]21)}
\nonumber\\
&&{}
-D_2^{([[42]5]31)}-D_2^{([[45]2]31)}
+D_2^{([[43]2]51)}-D_2^{([[42]3]51)}
+D_2^{([[13]5]24)}+D_2^{([[15]3]24)}
\nonumber\\
&&{}
-D_2^{([[12]5]34)}-D_2^{([[15]2]34)}
+D_2^{([[13]2]54)}-D_2^{([[12]3]54)}
\Big)
\label{eq:A-2comm}
\end{eqnarray}
and the terms with four commutators are
\begin{eqnarray}
\lefteqn{\frac{9}{64}(A_1^{(15(234))}-D_2^{(14(235))}-A_1^{(12345)}+D_2^{(12345)})_4}
\nonumber\\
&=&
-\frac{1}{5\cdot512}\Big(
\left(2A_1^{[[[[13]4]5]2]}+2A_1^{[[[[53]4]1]2]}+A_1^{[[[12]4][53]]}+A_1^{[[[52]4][13]]}+\mbox{Perm}(234)\right)
\nonumber\\
&&{}
+\left(2D_2^{[[[[13]2]4]5]}+2D_2^{[[[[43]2]1]5]}-D_2^{[[[15]2][43]]}+D_2^{[[[45]2][13]]}+\mbox{Perm}(235)\right)
\Big)\,.
\label{eq:A-4comm}
\end{eqnarray}
While the terms with three commutators are (with a bit of work using the Jacobi identity)
\begin{eqnarray}
\lefteqn{\frac{1}{64}(3K^{(16(3[24]5))}+2K^{(34([51]26))}+K^{(34([56]12))}+K^{([14]2356)}-K^{([56]1234)})_2}
\nonumber\\
&=&
\frac{1}{3\cdot256}\Big(
3K^{([14][23][56])}
+K^{([[13]5][24]6)}
-7K^{([[15]3][24]6)}
-4K^{([[51]3][46]2)}
-2K^{([[31]5][46]2)}
\nonumber\\
&&{}
-K^{([[63]5][14]2)}
-7K^{([[56]3][14]2)}
-7K^{([[13]2][56]4)}
+8K^{([[25]1][36]4)}
+8K^{([[15]6][23]4)}
\nonumber\\
&&{}
+2K^{([[56]1][23]4)}
+7K^{([[13]6][25]4)}
+4K^{([[36]1][25]4)}
+24K^{([[[14]2]3]56)}
+22K^{([[[14]3]2]56)}
\nonumber\\
&&{}
-K^{([[[13]4]2]56)}
+16K^{([[[61]3]5]24)}
+14K^{([[[63]1]5]24)}
+20K^{([[[61]5]3]24)}
+16K^{([[[65]1]3]24)}
\nonumber\\
&&{}
+18K^{([[[15]3]6]24)}
+15K^{([[[13]5]6]24)}
\Big)\,.
\label{eq:K-2comm}
\end{eqnarray}

With a bit more work using the Jacobi identity and the identities satisfied by $K$ the terms with five commutators can be written as
\begin{eqnarray}
\lefteqn{\frac{1}{64}(3K^{(16([24]35))}+2K^{(34([51]26))}+K^{(34([56]12))}+K^{([14]2356)}-K^{([56]1234)})_4}
\nonumber\\
&=&
-\frac{1}{90\cdot256}\Big(
-4K^{[[[[[14]3]2]5]6]}
-4K^{[[[[[14]6]3]2]5]}
+8K^{[[[[[46]3]1]5]2]}
+4K^{[[[[[36]4]1]5]2]}
\nonumber\\
&&{}
+K^{[[[[13]2]6][45]]}
+K^{[[[[13]6]2][45]]}
-4K^{[[[[14]2]3][56]]}
+35K^{[[[15]3][[24]6]]}
+17K^{[[[15]3][[46]2]]}
\nonumber\\
&&{}
-K^{[[[15]4][[23]6]]}
-9K^{[[[15]4][[36]2]]}
-24K^{[[[14]2][[56]3]]}
+6K^{[[[14]2][[36]5]]}
+54K^{[[[[36][15]]4]2]}
\nonumber\\
&&{}
+37K^{[[[[56][13]]4]2]}
-\frac{73}{2}K^{[[[[14][25]]3]6]}
+29K^{[[[15][24]][36]]}
-12K^{[[[14][56]][23]]}
+\frac{1}{2}K^{[[[16][45]][23]]}
\Big)\,.
\nonumber\\
\label{eq:K-4comm}
\end{eqnarray}
We now need to find what terms to include in $F_{0,2}^2$ in order to cancel these terms.

\subsection{Higher derivative/commutator terms}
In the abelian case there's a unique (up to field redefinitions) higher-derivative correction at order $\alpha'^4$, corresponding to the $\partial^4 F^4$-term in the action \cite{Collinucci:2002gd,Drummond:2004vf}. It can be written as a term
\begin{equation}
W=(\gamma^aD^cD^d\lambda)(\gamma^bD_cD_d\lambda)F_{ab}
\end{equation}
in $F_{\alpha\beta}$. In the abelian case this term is trivially closed (but not exact) in spinorial cohomology since $D_a$ reduces to $\partial_a$, so that it represents a supersymmetric correction to SYM, but in the non-abelian case a spinorial derivative of this term gives rise to commutator terms when the spinor derivative acts on the covariant derivatives. One can try to add other terms to cancel these extra commutator terms and working order by order in the number of commutators in the symmetrised expression. With a lot of work this can be done and we will now give the solution (some details are provided in Appendix C).


\subsubsection{The remaining terms in $F_{0,2}^2$}

The higher derivative/commutator terms in $F_{0,2}^2$ are
\begin{equation}
F^{HD1}=\frac{3}{256}(Q^{(123)}+\frac{1}{9}R^{(123)})\,,
\label{eq:FHD1}
\end{equation}
which cancels the terms with two and three commutators in $D_sF_{0,2}$ given in (\ref{eq:A-2comm}) and (\ref{eq:K-2comm}), and
\begin{eqnarray}
F^{HD2}&=&
\frac{1}{5\cdot256}\Big(2Q^{[1[23]]}-Q^{[[12]3]}
+\frac{1}{18}(R^{[1[23]]}-2R^{[[12]3]})
-12S+\frac{3}{4}T
\nonumber\\
&&{}
+\frac{5}{12}Z^{[[[16]2][[34]5]]}
-\frac{1}{3}Z^{[[[15]6][[23]4]]}
-\frac{1}{3}Z^{[[[63]1][[45]2]]}
-\frac{10}{3}Z^{[[[63][12]][45]]}
\Big)
\label{eq:FHD2}
\end{eqnarray}
which cancels the terms with four and five commutators given in (\ref{eq:A-4comm}) and (\ref{eq:K-4comm}). The expressions for $Q$, $R$, $S$ and $T$ are somewhat lengthy and are given in Appendix C which also contains some details of the calculation. The leading term in $Q^{(123)}$ gives precisely the abelian $\partial^4F^4$-correction mentioned above.

Thus we have completed the construction of $F_{0,2}$ at order $\alpha'^4$. It is given by the superembedding expression plus the higher derivative terms constructed above
\begin{equation}
F_{0,2}^2\sim9F^{SE}+F^{HD1}+F^{HD2}\,,
\end{equation}
where $F^{SE}=F^{redef}+F^{SE1}+F^{SE2}$ which are given in (\ref{eq:Fredef}), (\ref{eq:FSE1}) and (\ref{eq:FSE2}).


\section{Action form}


We recall that, given a closed superspace $D$-form, $L$, in a superspace with $D$ even dimensions, the integral of its purely even component over spacetime automatically defines a superinvariant \cite{Voronov,Gates:1997kr,Gates:1997ag}. Clearly, this integral is unchanged if an exact piece is added to $L$. In $D=10$ it turns out that the lowest non-trivial component of such a form is $L_{5,5}$, and that this will have the structure $L_{5,5}=\c_{5,2} M_{0,3}$. ($\c_{p,2}$ is the $p$-index symmetric gamma matrix considered as a $(p,2)$-form.) One reason for this is that the lowest component is $t_0$-closed and $H_t^{p,q}=0$ for $p>5$; the other is that $L$-forms with cohomologically non-trivial lowest components with $p<5$ do not contain singlet $L_{10,0}$ components. This fact sets up a direct correspondence between the ``ectoplasm'' approach and pure-spinor integration theory in $D=10$ \cite{Berkovits:2008qw}.

Furthermore, if we have a closed $(D+1)$-form $W=dZ$, where $Z$ is a potential $D$-form (Chern-Simons form), which is Weil trivial \cite{Bonora:1986xd}, i.e. $W=dK$ where $K$ is gauge-invariant, then we can easily construct a closed $D$-form $L=K-Z$. For $D=10$ SYM there are two natural, single-trace closed $11$-forms,

\bea
 W'&:=&\frac{1}{2} H_7 \Tr F^2 \nn\w1
 W&:=&\frac{1}{4} H_3 \Tr F^4\ ,
\la{6.1}
\eea

where $H_3$ and $H_7$ are closed forms which in flat superspace can be taken to be $\c_{1,2}$ and $\c_{5,2}$ respectively. At order $\a'^0$, $W'$ defines the on-shell standard SYM action while $W$ gives the $F^4$ invariant. However, the latter can also be described by $W'$ if we allow $\a'^2$ corrections to $F$ \cite{Berkovits:2008qw}. When we come to the $\a'^4$ corrections described above it turns out that both $W$s are required in order to achieve Weil triviality.

To see this, let us begin with the hypothesis that it is only necessary to use $W$. To describe the $\a'^4$ corrections we need to examine $W$ at first order ($\a'^2$), whereas we would need second order for $W'$. The lowest component of $W^1$ is

\be
 W^1_{4,7}=\c_{1,2} \symt((F^0_{1,1})^3 F^1_{0,2})\ .
\la{6.2}
\ee

Now we can write

\be
 \c_{1,2} F^0_{1,1}= t_0 X_{3,1}
\la{6.3}
\ee

where $X_{abc\d}=-\half(\c_{abc}\l)_\d$. Since both $F^0_{1,1}$ and $F^1_{0,2}$ are $t_0$-closed, it therefore follows that $W^1_{4,7}=t_0 K_{5,5}$ where $K^1_{5,5}=\symt (X_{3,1} (F^0_{1,1})^2 F^1_{0,2})$. If $W^1$ were to be exact by itself we would need to show that $W^1_{5,6}-d_1 K^1_{5,5}=t_0 K^1_{6,4}$ for some $K^1_{6,4}$. It is not difficult to see that this is not the case, and that we are left with

\bea
 W^1_{5,6}&-&d_1 K^1_{5,5} \sim -\symt(D_1X_{3,1} (F^0_{1,1})^2 F^1_{0,2}) +\gamma_{1,2}\symt(F^0_{2,0}(F^0_{1,1})^2F^1_{0,2})\nn\w1
 &\sim & \c_{5,2}\Tr (F^1_{0,2})^2\ ,
\la{6.4}
\eea

up to a constant. But the last term on the right is proportional to the $(5,6)$, i.e. lowest,  component of $W'^2$. So there is a linear combination, $\bbW$ say, of $W'$ and $\a'^2 W$ which is Weil-trivial up to order $\a'^4$. This follows because there are no further cohomological obstructions for $p>5$.

Although this is a nice approach to the action itself, it is not independent of the $F$ deformation calculation described above. This is because $\bbW$ must be closed, and since $W'$ must be considered up to $\a'^4$ this requires that $(D_1 F^1_{0,2})^2$ be trivial in $H_s^{0,3}$. Indeed, $F^2_{0,2}$ appears in $W'^2_{6,5}$ in the term $\c_{5,2} F^0_{1,1} F^2_{0,2}$. This is $t_0$ exact and therefore gives rise to a contribution to the $(7,3)$ component of $K$ which will in turn affect the action itself. (The action is given by $K_{10,0}$ as the Chern-Simons term has no purely bosonic part in flat superspace.)


\section{Conclusions}

We have extended the (unique) $\alpha'^2$-correction to ten-dimensional SYM, given by a symmetrised ordering of the corresponding abelian expression \cite{Cederwall:2001td}, to the next order, i.e. $\alpha'^4$. Until now there has not been a proof that this was possible, i.e. that the very non-trivial cohomological problem one has to solve has a solution (although string theory suggests that this should be the case). Moreover, we believe that the solution we have given is unique up to field redefinitions due to the absence of independent single-trace invariants at this order in $\a'$. In terms of the constraint on the dimension zero field strength (which defines SYM to a given order in $\alpha'$) we have found
\begin{equation}
F_{\alpha\beta}=F_{\alpha\beta}^{SE}
+\frac{3}{32}\sym((\gamma^aD^cD^d\lambda)_\a(\gamma^bD_cD_d\lambda)_\b F_{ab})
+\ldots
+\mathcal O(\alpha'^6)\,,
\end{equation}
where $F_{\alpha\beta}^{SE}$ is the result obtained from the generalised superembedding formalism \cite{Howe:2005jz,Howe:2006rv,Howe:2007eb} involving a symmetrised ordering of the fields. The second term reduces to the $\partial^4F^4$-correction in the abelian case and $\ldots$ denotes a number of terms involving commutators (with quite an intricate structure). This result has been established for the gauge group $U(k)$; it should be possible to consider others but not without significant changes to the details of the calculation.

We therefore see that the superembedding formalism with boundary fermions does not, by itself, seem to be able to give the complete correction at order $\alpha'^4$, and it is necessary to introduce further terms in order to accomplish this. Perhaps this situation could be remedied by an improved treatment of the quantum aspects of the boundary fermions. This is an interesting question which clearly requires a better understanding of the structure behind the complicated looking higher derivative/commutator terms if we are to make further progress.
Another interesting question is whether SYM at $\alpha'^4$ admits a second,
non-linearly realised, supersymmetry. This should be the case for there to be
an interpretation in terms of D9-branes in flat superspace.

We have worked here with the SYM-constraint rather than with the equations of motion or the action. In addition, in section 6, we saw that it is possible to construct the action from these results using the ``ectoplasm'' formalism, an interesting feature being the fact that eleven-forms involving both $\Tr F^2$ and $\Tr F^4$ terms are required.

Finally, we note that the situation may be different in $D<10$, because it may well be that the higher-derivative terms and some of the commutator terms could decouple. In the abelian case in $D=10$ the $F^4$ action can be derived from the eleven-form $W=H_3 F^4$, and this can be extended to $d^4 F^4$ by the insertion of derivatives in $W$. In the non-abelian case this is not possible because covariant derivatives would be required and their presence would spoil closure. This explains why these terms get induced at $\a'^4$ from the $\a'^2$ Born-Infeld deformation. However, in $D=9$ and below, the field strength superfield is a scalar, $\f_r,\, r=1,..n=(10-D),$ from which it is easy to construct $d^4 F^4$ invariants by integrating four powers of $\f$ over the full superspace. For $D\leq 8$ there are two such terms \cite{Drummond:2003ex}.

\vskip .5cm

{\bf Acknowledgements.}

The research of UL was supported by VR grant 621-2009-4066. LW was partially supported by INFN Special Initiative TV12. PSH and LW
thank the organisers of  the programme
``Geometrical aspects of String theory'' Nordita, Stockholm Oct. 15-
Dec. 15 2008'', where part of this work was done.
They would also like to thank the Department of Physics and Astronomy in Uppsala,
where part of this work was done, for hospitality and for providing a stimulating
atmosphere.

\newpage

\appendix


{\Large{\bf Appendices}}


\section{Identities involving symmetrised ordering}

The symmetrised product of $n$ fields $\Phi_1,\ldots,\Phi_n$ in the adjoint of the gauge-group $U(k)$ is defined as
\begin{equation}
\sym(\Phi_1\cdots\Phi_n)=\frac{1}{n!}\sum_{\substack{\{i_1,\ldots,i_n\}=\\\mbox{Perm}(1,\ldots,n)}}\Phi_{i_1}\cdots\Phi_{i_n}\,.
\end{equation}
Some useful identities are
\begin{eqnarray}
\sym(A\,\sym(BC))-\sym(ABC)&=&\frac{1}{12}([[A,B],C]+[[A,C],B])\\
\sym(A\,\sym(BCD))-\sym(ABCD)&=&\frac{1}{12}\sym([[A,B],C]D)+\mbox{Perm}(BCD)\\
\sym(A\,\sym(BCDE))-\sym(ABCDE)&=&\frac{1}{24}\sym([[A,B],C]DE)-\frac{1}{6!}[[[[A,B],C],D],E]\nonumber\\
&&{}+\mbox{Perm}(BCDE)\,.
\end{eqnarray}
We can now derive an identity that we will need. Taking $C$ to be $\sym(CDE)$ in the first identity above gives
\begin{eqnarray}
\sym(AB\,\sym(CDE))&=&\sym(A\,\sym(B\,\sym(CDE)))\nonumber\\
&&{}-\frac{1}{12}\left([[A,B],\sym(CDE)]+[[A,\sym(CDE)],B]\right)\,.
\end{eqnarray}
Using the other identities and the fact that the commutator acts as a derivation as well as the Jacobi identities this can be written
\begin{eqnarray}
\lefteqn{\sym(AB\,\sym(CDE))-\sym(ABCDE)}\nonumber\\
&=&\frac{1}{12}\left(\sym([[A,C],D]EB)+\sym([[B,C],D]EA)+\sym([A,C][B,D]E)\right)
\nonumber\\
&&{}-\frac{1}{360}\left(2[[[[A,C],B],D],E]+2[[[[B,C],A],D],E]+[[[A,E],D],[B,C]]+[[[B,E],D],[A,C]]\right)
\nonumber\\
&&{}+\mbox{Perm}(CDE)\,.
\end{eqnarray}
or more compactly, denoting the ordering of fields as a superscript,
\begin{eqnarray}
T^{(12(345))}-T^{(12345)}&=&\frac{1}{12}\left(T^{([[13]4]52)}+T^{([[23]4]51)}+T^{([13][24]5)}\right)
\nonumber\\
&&\hspace{-1cm}-\frac{1}{360}\left(2T^{[[[[13]4]2]5]}+2T^{[[[[23]4]1]5]}+T^{[[[13]4][25]]}+T^{[[[23]4][15]]}\right)
+\mbox{Perm}(345)\,.
\nonumber\\
\end{eqnarray}
This identity will be very useful for us since terms with this ordering appear in $D_sF_{0,2}$ at order $\alpha'^4$.


\section{Definitions of terms}


\subsection{Terms appearing in $(D_sF_{0,2})^2$}

The terms are defined for an arbitrary ordering of the fields. Spinor indices are suppressed and understood to be symmetrised (alternatively they can be thought of as contracted with spinorial vielbeins). All expressions are in $t_0$-cohomology and modulo the lowest order equations of motion (\ref{eq:Diraceqn}) and (\ref{eq:BIeqn}). The notation is explained further below (\ref{eq:DsF1}).
\begin{eqnarray}
A_1&=&\gamma^{bcdef}(\gamma^a\lambda)F_{cd}F_{ef}F_{bg}F^g{}_a\\
A_2&=&\gamma^{bcdef}(\gamma^a\lambda)F_{ab}F_{cd}F_{eg}F^g{}_f\\
A_3&=&\gamma^{cdefg}(\gamma^{ab}\gamma^h\lambda)F_{ab}F_{cd}F_{ef}F_{gh}
\end{eqnarray}
\begin{eqnarray}
B_1&=&i(\gamma^a\lambda)(\gamma^b\lambda)(\gamma^{cd}\gamma^e\lambda)D_bF_{cd}F_{ea}\\
B_2&=&i(\gamma^a\lambda)(\gamma^b\lambda)(\gamma^{cd}\gamma^e\lambda)F_{cd}D_bF_{ea}\\
B_3&=&i(\gamma^a\lambda)(\gamma^b\lambda)(\gamma^c\lambda)D_bF_{cd}F^d{}_a
\end{eqnarray}
\begin{eqnarray}
C_1&=&(\gamma^a\lambda)(\gamma^b\lambda)(\gamma^cD_b\lambda)\lambda\gamma_cD_a\lambda\\
C_2&=&(\gamma^a\lambda)(\gamma^b\lambda)(\gamma^cD_b\lambda)\lambda\gamma_aD_c\lambda
\end{eqnarray}
\begin{eqnarray}
D_1&=&i(\gamma^a\lambda)(\gamma^{cd}\gamma^b\lambda)(\gamma^eD_a\lambda)F_{cd}F_{be}\\
D_2&=&i(\gamma^a\lambda)(\gamma^{cd}\gamma^e\lambda)(\gamma^bD_e\lambda)F_{ab}F_{cd}\\
D_3&=&i(\gamma^a\lambda)(\gamma^{bcd}\lambda)(\gamma^eD_b\lambda)F_{ac}F_{de}\\
D_4&=&i(\gamma^a\lambda)(\gamma^b\lambda)(\gamma^{cd}\gamma^eD_b\lambda)F_{cd}F_{ea}\\
D_5&=&i(\gamma^a\lambda)(\gamma^b\lambda)(\gamma^cD^d\lambda)F_{bc}F_{da}\\
D_6&=&i(\gamma^a\lambda)(\gamma^b\lambda)(\gamma^cD^d\lambda)F_{ab}F_{cd}\\
D_7&=&i(\gamma^a\lambda)(\gamma^b\lambda)(\gamma^cD_b\lambda)F_{cd}F^d{}_a
\end{eqnarray}
\begin{eqnarray}
E&=&(\gamma^a\lambda)(\gamma^b\lambda)(\gamma^c\lambda)\lambda\gamma_aD_bD_c\lambda
\end{eqnarray}
\begin{eqnarray}
G&=&\gamma^{cdefg}(\gamma^bD^a\lambda)D_aF_{cd}F_{bg}F_{ef}
\end{eqnarray}
\begin{eqnarray}
H&=&i(\gamma^a\lambda)(\gamma^bD_c\lambda)(\gamma^cD^d\lambda)D_dF_{ab}
\end{eqnarray}
\begin{eqnarray}
I_1&=&i(\gamma^a\lambda)(\gamma^bD^c\lambda)(\gamma^dD_cD_a\lambda)F_{bd}\\
I_2&=&i(\gamma^a\lambda)(\gamma^bD^c\lambda)(\gamma^dD_cD_b\lambda)F_{ad}
\end{eqnarray}
\begin{eqnarray}
J&=&i(\gamma^aD^b\lambda)(\gamma^cD_b\lambda)(\gamma^dD_a\lambda)F_{cd}
\end{eqnarray}
\begin{eqnarray}
K&=&(\gamma^a\lambda)(\gamma^b\lambda)(\gamma^c\lambda)\lambda\gamma_c\lambda F_{ab}\,.
\end{eqnarray}

\subsubsection{Some important identities involving these terms}

Besides trivial identities such as $A_1^{12345}=A_1^{13245}$ or $A_2^{12345}=-A_2^{12354}$ some slightly more non-trivial identities are
\begin{equation}
E^{12345}=E^{13245}-K^{3124[56]}
\end{equation}
and
\begin{eqnarray}
K^{123456}&=&-K^{125436}-K^{124356}\,.
\end{eqnarray}
Making use of the gamma-matrix identity
\begin{equation}
\gamma^{[abcde}(\gamma^{f]}\cdots)\sim0\,,
\end{equation}
we find some further useful identities
\begin{eqnarray}
A_2^{12345}&\sim&-A_2^{13245}+\frac{1}{2}A_1^{12345}-\frac{1}{2}A_1^{12354}\\
A_3^{12534}&\sim&-A_3^{12354}-A_3^{12345}-4A_1^{15423}-4A_1^{15234}-4A_1^{15243}
\end{eqnarray}
and finally

\begin{eqnarray}
\lefteqn{i\gamma^{bcdef}(\gamma^a\lambda)\lambda\gamma_aD_b\lambda F_{cd}F_{ef}}
\nonumber\\
&\sim&
-4D_1^{12345}-4D_1^{12354}
+4D_2^{12345}+4D_2^{12354}
+8D_3^{12345}+8D_3^{12354}
+4D_4^{12345}+4D_4^{12354}
\nonumber\\
&&{}
-4D_4^{21345}-4D_4^{21354}
+8D_5^{12345}+8D_5^{12354}
-8D_6^{12345}-8D_6^{12354}
-8D_7^{12345}-8D_7^{12354}
\nonumber\\
&&{}
+16D_7^{21345}+16D_7^{21354}\,.
\end{eqnarray}


\subsection{Terms appearing in $F_{0,2}^2$}

The terms we will need consist of terms with three to six fields. We define

\subsubsection{Three-field terms}

\begin{eqnarray}
W=(\gamma^aD^cD^d\lambda)(\gamma^bD_cD_d\lambda)F_{ab}
\end{eqnarray}

\subsubsection{Four-field terms}

$D\lambda D\lambda FF$:
\begin{eqnarray}
X_1&=&(\gamma^aD^b\lambda)(\gamma_cD_b\lambda)F_{ad}F^d{}_c\\
X_2&=&(\gamma^aD^c\lambda)(\gamma^bD^d\lambda)F_{ab}F_{cd}\\
X_3&=&(\gamma^{ab}\gamma^cD_e\lambda)(\gamma^dD^e\lambda)F_{ab}F_{cd}
\end{eqnarray}
$D\lambda\lambda DFF$:
\begin{eqnarray}
X_4&=&(\gamma^aD^b\lambda)(\gamma^c\lambda)D_bF_{cd}F^d{}_a\\
X_5&=&(\gamma^aD^b\lambda)(\gamma^{cd}\gamma^e\lambda)D_bF_{ae}F_{cd}
\end{eqnarray}
$D\lambda D\lambda D\lambda\lambda$:
\begin{eqnarray}
X_6&=&i(\gamma^aD^b\lambda)(\gamma^cD_a\lambda)\lambda\gamma_cD_b\lambda
\end{eqnarray}


\subsubsection{Five-field terms}

$\lambda^2F^3$:
\begin{eqnarray}
Y_1&=&(\gamma^a\lambda)(\gamma^b\lambda)F_{ab}F_{cd}F^{cd}\\
Y_2&=&(\gamma^a\lambda)(\gamma^d\lambda)F_{ab}F^{bc}F_{cd}\\
Y_3&=&(\gamma^{ab}\gamma^c\lambda)(\gamma^d\lambda)F_{ab}F_{ce}F^e{}_d\\
Y_4&=&(\gamma^{ab}\gamma^c\lambda)(\gamma^{ef}\gamma^d\lambda)F_{ab}F_{cd}F_{ef}\\
Y_5&=&(\gamma^a\gamma^{cdef}\lambda)(\gamma^b\lambda)F_{ab}F_{cd}F_{ef}
\end{eqnarray}

$\lambda^3D\lambda F$:
\begin{eqnarray}
Y_6&=&i(\gamma^a\lambda)(\gamma^bD_c\lambda)\lambda\gamma^c\lambda F_{ab}\\
Y_7&=&i(\gamma^a\lambda)(\gamma^b\lambda)\lambda\gamma^cD_a\lambda F_{bc}\\
Y_8&=&i(\gamma^a\lambda)(\gamma^b\lambda)\lambda\gamma_aD^c\lambda F_{bc}\\
Y_9&=&i(\gamma^a\lambda)(\gamma^{cd}\gamma^b\lambda)\lambda\gamma_aD_b\lambda F_{cd}\\
Y_{10}&=&i(\gamma^a\lambda)(\gamma^b\gamma^{cd}\lambda)\lambda\gamma_bD_a\lambda F_{cd}\\
Y_{11}&=&i(\gamma^a\lambda)(\gamma^b\lambda)\lambda\gamma^{cd}\gamma_bD_a\lambda F_{cd}\\
Y_{12}&=&i(\gamma^a\lambda)(\gamma^b\gamma^{cd}\lambda)\lambda\gamma_cD_d\lambda F_{ab}
\end{eqnarray}

$\lambda^4DF$:
\begin{eqnarray}
Y_{13}&=&i(\gamma^a\lambda)(\gamma^b\lambda)\lambda\gamma_a{}^{cd}\lambda\,D_bF_{cd}
\end{eqnarray}

\subsubsection{Six-field terms}
\begin{eqnarray}
Z&=&(\gamma^a\lambda)(\gamma^b\lambda)\lambda\gamma_a\lambda \lambda\gamma_b\lambda
\end{eqnarray}

\subsection{The spinorial derivative of these terms}
For our computations we need to know the spinorial derivative of these terms (modulo lowest order equations of motion). We compute them for an arbitrary ordering of the fields. We find
\begin{eqnarray}
2D_sW\sim D_5^{[13][52]4}+D_5^{[52][13]4}+2I_1^{[12]34}+2I_1^{3[12]4}+I_2^{[12]34}+I_2^{3[12]4}
\end{eqnarray}
\begin{eqnarray}
2D_sX_1&\sim&D_7^{[12]354}-D_7^{3[12]45}-J^{1234}-J^{2143}\\
2D_sX_2&\sim&-D_5^{[12]345}+D_5^{3[12]45}+J^{2341}+J^{3241}\\
2D_sX_3&\sim&D_1^{[12]345}-D_4^{3[12]45}+G^{2143}-4J^{2143}\\
2D_sX_4&\sim&B_3^{[12]345}+D_7^{31[25]4}-H^{3142}-I_1^{2134}\\
2D_sX_5&\sim&-B_2^{[12]354}-D_1^{32[15]4}-G^{1324}-G^{1423}+4I_2^{2134}\\
2D_sX_6&\sim&-C_1^{[12]345}-C_2^{354[12]}-C_2^{351[42]}-4H^{3214}+4J^{1342}
\end{eqnarray}
\begin{eqnarray}
D_sY_1&\sim&-D_6^{12435}-D_6^{12453}\\
2D_sY_2&\sim&D_5^{12534}-D_5^{21435}+D_7^{21345}-D_7^{12543}\\
2D_sY_3&\sim&-A_1^{21345}-D_1^{21453}-D_2^{21534}-4D_5^{21435}\\
2D_sY_4&\sim&-A_3^{31452}-A_3^{13254}+4D_2^{21534}-4D_2^{12435}\\
D_sY_5&\sim&A_1^{41523}+A_1^{21543}+A_1^{21534}+2A_2^{51234}+2A_2^{51243}+2D_2^{21435}+2D_2^{21453}+2D_3^{21435}
\nonumber\\
&&{}
+2D_3^{21453}+4D_5^{12453}+4D_5^{12435}+2D_5^{21534}+2D_5^{21543}+2D_7^{21543}+2D_7^{21534}
\end{eqnarray}
\begin{eqnarray}
2D_sY_6&\sim&-D_2^{13254}+D_2^{13524}+K^{1[23]456}\\
2D_sY_7&\sim&B_1^{21345}-C_1^{12453}-D_4^{21435}\\
2D_sY_8&\sim&4B_3^{21345}-4B_3^{31245}-C_2^{12453}+4D_5^{21435}+K^{314[52]6}\\
D_sY_9&\sim&-2B_2^{13254}+2B_2^{23154}-2D_1^{14235}-2D_1^{15234}-2D_2^{12435}+2D_2^{14235}
\nonumber\\
&&{}
+2D_2^{15234}+4D_3^{14235}+4D_3^{15234}+2D_4^{14235}+2D_4^{15234}-2D_4^{24135}-2D_4^{25134}
\nonumber\\
&&{}
+4D_5^{14235}+4D_5^{15234}-4D_6^{14235}-4D_6^{15234}-4D_7^{14235}-4D_7^{15234}+8D_7^{24135}
\nonumber\\
&&{}
+8D_7^{25134}+2K^{314[52]6}\\
D_sY_{10}&\sim&2B_2^{23154}-8B_3^{21345}+2C_1^{12453}-2D_1^{12534}-2D_1^{14235}-2D_1^{15234}
\nonumber\\
&&{}
-2D_4^{24135}-2D_4^{25134}+8D_7^{21435}+8D_7^{24135}+8D_7^{25134}
\\
D_sY_{11}&\sim&2B_2^{21354}-8B_3^{23145}+2C_1^{14253}+2C_1^{24135}-2C_1^{21435}+2C_2^{21435}+2D_4^{21534}
\nonumber\\
&&{}
+2D_4^{21435}-8D_7^{21534}
\\
D_sY_{12}&\sim&B_1^{12345}-B_1^{13245}-4B_3^{12345}+4B_3^{13245}-D_2^{15234}-2D_3^{15234}-2D_3^{12534}-D_4^{14235}
\nonumber\\
&&{}
-D_4^{12435}-2D_5^{12435}-2D_5^{14235}+2D_6^{15234}+2D_6^{12534}+2D_7^{12435}+2D_7^{14235}
\nonumber\\
&&{}
-K^{134[52]6}
\end{eqnarray}
\begin{eqnarray}
D_sY_{13}&\sim&2B_1^{12354}+2B_1^{12534}+B_2^{12435}+B_2^{12345}-8B_3^{12354}-8B_3^{12534}-2E^{12345}
\nonumber\\
&&{}
+2E^{12435}+E^{32145}-E^{32415}
\end{eqnarray}
and
\begin{eqnarray}
D_sZ&\sim&-2K^{136245}+2K^{132645}+2K^{324561}-2K^{324516}\,.
\end{eqnarray}


\section{Higher derivative/commutator terms}


In this appendix we will construct the combinations of terms from Appendix B which are needed to write $F_{0,2}$ at order $\alpha'^4$. Let us define
\begin{eqnarray}
Q^{123}&=&
8W^{123}
-48X_1^{[14]23}
+16X_1^{[31]24}
+8X_1^{[34]21}
+24X_2^{[12]43}
+16X_2^{[13]42}
-4X_3^{[42]13}
\nonumber\\
&&{}
-4X_3^{[32]14}
+32X_4^{[12]34}
-4X_5^{[12]34}
+8X_6^{[13]42}
\nonumber\\
&&{}
-24Y_2^{[35][12]4}
+16Y_2^{[23][15]4}
-16Y_2^{[34][12]5}
+16Y_2^{[[24]1]35}
\nonumber\\
&&{}
-4Y_3^{[53][12]4}
+4Y_3^{[51][24]3}
-4Y_3^{[54][12]3}
-4Y_3^{[[12]4]35}
-4Y_3^{[[12]4]53}
\nonumber\\
&&{}
-4Y_3^{[[24]1]53}
-4Y_3^{[[24]1]35}
-4Y_3^{[[12]3]54}
-4Y_3^{[[23]1]45}
\nonumber\\
&&{}
+Y_4^{[[15]4]23}
-Y_4^{[[12]4]53}
+Y_4^{[[15]3]24}
-Y_4^{[[12]3]54}
-Y_4^{[[54]1]23}
-Y_4^{[[24]1]53}
\nonumber\\
&&{}
+2Y_6^{[35][14]2}
-2Y_6^{[53][12]4}
+2Y_6^{[52][13]4}
+Y_6^{[[34]5]12}
-2Y_6^{[[52]3]14}
\nonumber\\
&&{}
+2Y_6^{[[53]2]14}
+2Y_6^{[[12]3]45}
+10Y_6^{[[13]2]45}
+2Y_6^{[[13]4]25}
-4Y_6^{[[13]5]42}
\nonumber\\
&&{}
+4Y_8^{[[23]1]45}
-8Y_8^{[35][12]4}
-8Y_8^{[15][23]4}
-Y_{10}^{[[13]2]45}
+Y_{10}^{[[12]3]54}
-Y_{11}^{[[13]2]54}
\end{eqnarray}
and
\begin{eqnarray}
R^{123}&=&
-12Z^{[61][23][45]}
-4Z^{[[15]4][23]6}
+Z^{[[14]5][23]6}
+Z^{[[12]3][45]6}
+2Z^{[[13]2][45]6}
\nonumber\\
&&{}
-4Z^{[[[15]3]2]46}
+Z^{[[[13]5]2]46}
-2Z^{[[[15]2]3]46}
+2Z^{[[[25]1]3]46}\,.
\end{eqnarray}
Using the spinorial derivatives computed in Appendix B we find
\begin{eqnarray}
D_sQ^{123}&\sim&
8I_1^{3[12]4}
-8I_1^{[12]34}
+4I_2^{3[12]4}
-4I_2^{[12]34}
-4J^{[34]12}
-4J^{[34]21}
-2A_1^{[53][12]4}
\nonumber\\
&&{}
-2A_1^{[52][14]3}
-2A_1^{[54][12]3}
-2A_1^{[[12]4]35}
-2A_1^{[[14]2]35}
-2A_1^{[[13]2]45}
-2A_1^{[[54]2]13}
\nonumber\\
&&{}
-2A_1^{[[52]3]14}
-2A_1^{[[52]4]13}
-D_2^{[45][13]2}
+D_2^{[43][15]2}
+D_2^{[45][12]3}
-D_2^{[42][15]3}
\nonumber\\
&&{}
-D_2^{[42][13]5}
+D_2^{[43][12]5}
+D_2^{[[43]5]12}
+D_2^{[[45]3]12}
-D_2^{[[43]2]15}
+D_2^{[[42]3]15}
\nonumber\\
&&{}
-D_2^{[[42]5]13}
-D_2^{[[45]2]13}
-D_2^{[[15]3]24}
-D_2^{[[13]5]24}
+D_2^{[[12]5]34}
+D_2^{[[15]2]34}
\nonumber\\
&&{}
+D_2^{[[12]3]54}
-D_2^{[[13]2]54}
-8D_5^{[25][14]3}
-8D_5^{[14][25]3}
+4D_5^{[13][52]4}
-4D_5^{[52][13]4}
\nonumber\\
&&{}
-8D_5^{[34][12]5}
+8D_5^{[34]5[12]}
-8D_5^{[[12]3]54}
+8D_5^{[[12]3]45}
+8D_7^{[13][25]4}
\nonumber\\
&&{}
-8D_7^{[25][13]4}
+4D_7^{[54][12]3}
-4D_7^{[54]3[12]}
-8D_7^{[[12]4]35}
+8D_7^{[[12]4]53}
\nonumber\\
&&{}
+5K^{[46][15][23]}
-K^{[[34]6]1[25]}
-K^{[[13]4][25]6}
+K^{[[14]3][25]6}
+K^{[64][[13]2]5}
\nonumber\\
&&{}
-K^{[[13]6][24]5}
-2K^{[[14]6]5[23]}
+K^{[[[13]6]4]25}
-K^{[[64][13]]25}
+K^{[[[13]2]4]56}
\nonumber\\
&&{}
-5K^{[[14][23]]56}
\end{eqnarray}
and
\begin{eqnarray}
D_sR^{123}&\sim&
-24K^{[15][36][24]}
+24K^{[15][32][64]}
+24K^{[13][24][56]}
-24K^{[63][24][15]}
+8K^{[[13]5][24]6}
\nonumber\\
&&{}
-2K^{[[35]1][24]6}
-2K^{[[32]4][51]6}
-4K^{[[34]2][51]6}
-8K^{[[36]5][14]2}
-2K^{[[35]6][14]2}
\nonumber\\
&&{}
-4K^{[[34]1][56]2}
-2K^{[[31]4][56]2}
+8K^{[[14]2][36]5}
+2K^{[[13]2][64]5}
+8K^{[[14]6][32]5}
\nonumber\\
&&{}
+2K^{[[16]4][32]5}
-2K^{[[13]6][24]5}
+4K^{[[16]3][24]5}
+8K^{[[[14]2]3]65}
+4K^{[[[14]3]2]65}
\nonumber\\
&&{}
-4K^{[[[34]1]2]65}
-2K^{[[[34]1]2]56}
+8K^{[[[31]4]2]56}
+4K^{[[[31]2]4]56}
+8K^{[[[14]6]3]25}
\nonumber\\
&&{}
+2K^{[[[16]4]3]25}
-4K^{[[[14]3]6]25}
+4K^{[[[34]1]6]25}
-8K^{[[[36]4]1]52}
-2K^{[[[34]6]1]52}
\nonumber\\
&&{}
-4K^{[[[36]1]4]52}
+4K^{[[[16]3]4]52}\,.
\end{eqnarray}
This is enough to account for the terms with two or three commutators in $F_{0,2}^2$, (\ref{eq:FHD1}).

To account for the terms with four and five commutators we can consider
\begin{eqnarray}
\lefteqn{2D_sQ^{[1[23]]}-D_sQ^{[[12]3]}}
\nonumber\\
&\sim&
12J^{[[12][34]]}
+4A_1^{[[[12]4][53]]}
+4A_1^{[[[14]2][53]]}
+4A_1^{[[[12]3][54]]}
-2A_1^{[[[54]3][12]]}
-2A_1^{[[[52]3][14]]}
\nonumber\\
&&{}
-2A_1^{[[[52]4][13]]}
+2A_1^{[[[[12]4]5]3]}
+2A_1^{[[[[14]2]5]3]}
+2A_1^{[[[[13]2]5]4]}
+2A_1^{[[[[54]2]1]3]}
+2A_1^{[[[[52]3]1]4]}
\nonumber\\
&&{}
+2A_1^{[[[[52]4]1]3]}
-2D_2^{[[[12]3][45]]}
+2D_2^{[[[15]3][42]]}
+2D_2^{[[[13]5][42]]}
+2D_2^{[[[13]2][45]]}
-2D_2^{[[[12]5][43]]}
\nonumber\\
&&{}
-2D_2^{[[[15]2][43]]}
+D_2^{[[[43]5][12]]}
+D_2^{[[[45]3][12]]}
-D_2^{[[[43]2][15]]}
+D_2^{[[[42]3][15]]}
-D_2^{[[[42]5][13]]}
\nonumber\\
&&{}
-D_2^{[[[45]2][13]]}
-D_2^{[[[[43]5]1]2]}
-D_2^{[[[[45]3]1]2]}
+D_2^{[[[[43]2]1]5]}
-D_2^{[[[[42]3]1]5]}
+D_2^{[[[[42]5]1]3]}
\nonumber\\
&&{}
+D_2^{[[[[45]2]1]3]}
+D_2^{[[[[15]3]4]2]}
+D_2^{[[[[13]5]4]2]}
-D_2^{[[[[12]5]4]3]}
-D_2^{[[[[15]2]4]3]}
-D_2^{[[[[12]3]4]5]}
\nonumber\\
&&{}
+D_2^{[[[[13]2]4]5]}
+24D_5^{[[[12]5][34]]}
-24D_5^{[[[12]3][54]]}
+12D_7^{[[[12]3][45]]}
+24D_7^{[[[12]4][53]]}
\nonumber\\
&&{}
+\frac{1}{2}K^{[[[[13]6]2][45]]}
-\frac{1}{2}K^{[[[[13]2]6][45]]}
+3K^{[[[[13]2]4][56]]}
+2K^{[[[15]3][[24]6]]}
-5K^{[[[15]4][[23]6]]}
\nonumber\\
&&{}
+2K^{[[[14]2][[36]5]]}
+2K^{[[[14]2][[56]3]]}
-\frac{9}{2}K^{[[[[14][25]]3]6]}
-K^{[[[[13][56]]4]2]}
+\frac{23}{2}K^{[[[14][25]][36]]}
\nonumber\\
&&{}
+6K^{[[[15][46]][23]]}
-\frac{1}{2}K^{[[[16][45]][23]]}\,.
\label{eq:Q123comm}
\end{eqnarray}
This is almost what we need but a few terms need to be added to the LHS to cancel some terms. To cancel the remaining $J$-term we use the fact that
\begin{eqnarray}
D_s(2X_2^{4123}-2X_1^{1324}+2X_2^{1342})
&\sim&
J^{1234}-J^{3412}
-D_5^{5[12]34}+D_5^{53[12]4}
\nonumber\\
&&{}
-D_5^{[12]453}+D_5^{345[12]}
-D_7^{[12]534}+D_7^{34[12]5}\,,
\end{eqnarray}
so that defining
\begin{equation}
S=X_2^{[[41][23]]}-X_1^{[[13][24]]}+X_2^{[[13][42]]}+2Y_2^{[[[12]3][45]]}-2Y_2^{[[[12]4][35]]}
\end{equation}
we get
\begin{eqnarray}
D_sS&\sim&J^{[[12][34]]}+2D_5^{[[[12]5][34]]}-2D_5^{[[[12]3][54]]}+D_7^{[[[21]3][45]]}+2D_7^{[[[12]4][53]]}\,.
\end{eqnarray}
This takes care of the $J$, $D_5$ and $D_7$-terms on the RHS of (\ref{eq:Q123comm}).

To cancel the last unwanted terms we use the fact that
\begin{eqnarray}
\lefteqn{D_s(-Y_4^{12435}+2Y_4^{24351}+4Y_4^{34512}-2Y_4^{23541}-2Y_4^{12345})}
\nonumber\\
&\sim&
-\frac{1}{2}(-A_3^{31542}-A_3^{13524}+4D_2^{21354}-4D_2^{12345})
+A_3^{15423}+A_3^{35241}-4D_2^{13542}+4D_2^{23451}
\nonumber\\
&&{}
+2(-A_3^{45231}-A_3^{25413}+4D_2^{53421}-4D_2^{43512})
-A_3^{14253}-A_3^{32451}+4D_2^{15432}-4D_2^{24531}
\nonumber\\
&&{}
+A_3^{31452}+A_3^{13254}-4D_2^{21534}+4D_2^{12435}\,.
\end{eqnarray}
Defining
\begin{eqnarray}
T&=&
-Y_4^{[[[12]4][35]]}
+2Y_4^{[[[24]3][51]]}
+4Y_4^{[[[34]5][12]]}
-2Y_4^{[[[23]5][41]]}
-2Y_4^{[[[12]3][45]]}
\nonumber\\
&&{}
+8Y_6^{[[[14]5][23]]}
-12Y_6^{[[[45]2][13]]}
-4Y_6^{[[[13]2][45]]}
-4Y_6^{[[[12]3][45]]}
-4Y_6^{[[[13]4][52]]}
\nonumber\\
&&{}
+4Y_6^{[[[52]3][41]]}
+4Y_6^{[[[53]4][12]]}
\end{eqnarray}
then gives, using the identity satisfied by $A_3$,
\begin{eqnarray}
D_sT&\sim&
-4A_1^{[[[14]2][53]]}
-4A_1^{[[[12]3][54]]}
-4A_1^{[[[12]4][53]]}
+4A_1^{[[[54]2][13]]}
+4A_1^{[[[52]3][14]]}
\nonumber\\
&&{}
+4A_1^{[[[52]4][13]]}
+2D_2^{[[[12]3][45]]}
-2D_2^{[[[13]2][45]]}
-2D_2^{[[[13]5][42]]}
-2D_2^{[[[15]3][42]]}
\nonumber\\
&&{}
+2D_2^{[[[12]5][43]]}
+2D_2^{[[[15]2][43]]}
-2D_2^{[[[42]3][15]]}
+2D_2^{[[[43]2][15]]}
-2D_2^{[[[43]5][12]]}
\nonumber\\
&&{}
-2D_2^{[[[45]3][12]]}
+2D_2^{[[[42]5][13]]}
+2D_2^{[[[45]2][13]]}
-4K^{[[[[13]2]4][56]]}
+2K^{[[[15]3][[24]6]]}
\nonumber\\
&&{}
+2K^{[[[15]3][[46]2]]}
+6K^{[[[15]4][[23]6]]}
+2K^{[[[14]5][[36]2]]}
-4K^{[[[14]2][[36]5]]}
\nonumber\\
&&{}
-4K^{[[[14]2][[56]3]]}
-2K^{[[[[13][56]]4]2]}
+K^{[[[[14][25]]3]6]}
+4K^{[[[14][56]][23]]}
+K^{[[[16][45]][23]]}\,,
\nonumber\\
\end{eqnarray}
where the $K$-terms have been simplified. This gives the last terms we need to construct $F_{0,2}$ at order $\alpha'^4$, (\ref{eq:FHD2}).

\end{document}